\documentclass[preprintnumbers,eqsecnum,amsmath,amssymb,twocolumn,nofootinbib]{revtex4}

\usepackage{graphicx}
\usepackage{dcolumn}
\usepackage{bm}
\usepackage{epsfig}

\newcommand{\cc}{\centering}

\newcommand{\be}{\begin{eqnarray}}
\newcommand{\ee}{\end{eqnarray}}
\newcommand{\<}{\langle}
\renewcommand{\>}{\rangle}

\newcommand{\ve}{\varepsilon}
\newcommand{\mc}{\mathcal}
\newcommand{\mbf}{\mathbf}
\newcommand{\nn}{\nonumber}
\newcommand{\NP}[4]{{#1}, Nucl. Phys. \textbf{#2}, {#3} ({#4})}
\newcommand{\NPps}[4]{{#1}, Nucl. Phys. Proc. Suppl. \textbf{#2}, {#3} ({#4})}
\newcommand{\PPNP}[4]{{#1}, Prog. Part. Nucl. Phys. \textbf{#2}, {#3} ({#4})}

\newcommand{\PL}[4]{{#1}, Phys. Lett. \textbf{#2}, {#3} ({#4})}
\newcommand{\PR}[4]{{#1}, Phys. Rev. \textbf{#2}, {#3} ({#4})}

\begin{document}

\preprint{DESY 04-159}
\preprint{RM3-TH/04-20}
\preprint{ROME1-1387/2004}

\title{Extracting excited states from lattice QCD: the Roper resonance\footnote{\bf To appear in Physics Letters B.}}

\author{D.~Guadagnoli$^a$, M.~Papinutto$^b$ and S.~Simula$^c$}

\affiliation{ \vspace{0.5cm}
$^a$Dipartimento di Fisica, Universit\`a di Roma ``La Sapienza'', 
    and INFN, Sezione di Roma, P.le A.~Moro 2, I-00185 Rome, Italy\\
$^b$NIC/DESY Zeuthen, Platanenallee 6, D-15738 Zeuthen, Germany\\
$^c$INFN, Sezione di Roma Tre, Via della Vasca Navale 84, I-00146, Roma, 
Italy}

\begin{abstract}
We present a new method for extracting excited states from a single two-point
correlation function calculated on the lattice. Our method simply combines the 
correlation function evaluated at different time slices so as to ``subtract'' 
the leading exponential decay (ground state) and to give access to the first excited 
state. The method is applied to a quenched lattice study (volume = $24^3 \times 64$, 
$\beta = 6.2$, $a^{-1} = 2.55$ GeV) of the first excited state of the nucleon 
using the local interpolating operator $\mc{O} = \ve_{abc}[u^T_a C \gamma_5 d_b]u_c$.
The results are consistent with the identification of our extracted excited 
state with the Roper resonance $N'(1440)$. The switching of the level ordering 
with respect to the negative-parity partner of the nucleon, $N^*(1535)$, is 
not seen at the simulated quark masses and, basing on crude extrapolations, is 
tentatively expected to occur close to the physical point.
\end{abstract}

\maketitle

\section{\protect Introduction \label{sec:introduction}}

Hadron spectroscopy offers a well defined framework to test the
non-perturbative regime of QCD and to shed light on the mechanism responsible
for confinement. Of course, the masses of ground states do not offer sufficient 
information to unravel the dynamics responsible for that mechanism, but fundamental
insight can be gained by accessing the masses of excited states as well.

As is well known, one of the long-standing puzzles in baryon spectroscopy is
the level ordering between the negative-parity partner of the nucleon,
$N^*(1535)$, and the first excitation of the nucleon, the so-called Roper resonance 
$N'(1440)$. Phenomenological low-energy models of QCD are generally quite successful 
in explaining many features of baryon mass spectra \cite{Capstick}. Some of them are 
however unable to explain the proper level ordering between positive- and negative-parity 
excited states of the nucleon, like e.g.~the one-gluon-exchange quark model of 
Ref.~\cite{Isgur}; such a failure has led to the speculation that the Roper 
resonance cannot be described by a picture based only on three constituent quarks.
Other models do reproduce the correct level ordering, like the Goldstone-boson-exchange 
quark model of Ref.~\cite{Glozman}, where the level switching between the radial and orbital
excitations of the nucleon is a consequence of the spontaneous breaking of chiral 
symmetry.

It is clear that a non-perturbative approach based on first principles, such as lattice QCD, 
offers a unique possibility to unfold the above issues and, in particular, to 
investigate the evolution of the excitations of the nucleon as a function of the quark masses.
Ground-state spectroscopy in lattice QCD is well established and there is a general agreement 
with experimental data at the level of $5 \div 10 \%$ precision \cite{ground}. On the contrary
it is still challenging to measure the mass of an excited state. Both ground and excited 
masses are extracted from the (Euclidean) time behavior of two-point correlators and therefore 
a clean separation among them is required. This problem is currently addressed following two 
main approaches\footnote{We mention that a throughout discussion of various mathematical 
techniques for analyzing exponentially-damped time series can be found in Ref.~\cite{fleming}.}.

The first approach uses a single two-point correlator, constructed from an
interpolating field with appropriate quantum numbers. A naive multiple-exponential
fit of the time dependence of the latter is in general affected by large systematic 
or statistical errors, due to the contamination from high excited states and/or 
to the large number of fitting parameters. To circumvent this problem, 
methods inspired by Bayesian statistics, like the constrained curve fitting method 
\cite{lepage} or the maximum entropy method \cite{mem}, were developed and 
applied to lattice analyses aimed at studying the Roper resonance \cite{dong,sasaki}. 
The crucial point of these methods is clearly the choice of the priors (see 
Ref.~\cite{priors}).

The second approach uses several two-point correlators constructed from various 
interpolating operators $\mc{O}_i$ which couple differently to the states of interest 
\cite{variational}. A matrix of correlation functions is constructed choosing in all 
possible (independent) ways the source and the sink operators. By solving an appropriate 
eigenvalue problem, the contributions coming from the various states coupled to the operators 
$\mc{O}_i$ are disentangled, so that the masses of ground and excited states can be 
determined from the eigenvalues (see Refs.~\cite{UKQCD,leinweber}). The interpolating 
operators must be chosen properly, taking into account the computational cost of 
constructing a matrix of correlators. The following two (local) operators are commonly 
employed \cite{leinweber,sbo,bgr1}:
\be
&&\mc{O}(x) = \ve_{abc}[u^T_a C \gamma_5 d_b]u_c(x)~,\nn \\
&&\mc{O}'(x) = \ve_{abc}[u^T_a C d_b] \gamma_5 u_c(x)~,
\label{ops}
\ee
where $C$ is the charge conjugation Dirac matrix. It is known \cite{richards} that the 
first and the second operators in Eq.~(\ref{ops}) are, respectively, strongly and very 
weakly coupled to the ground state of the nucleon. A recent improvement \cite{bgr2} is 
the use of different smeared interpolating operators as a tool to generate nodes in the 
radial wave function, which allows to capture more efficiently the Roper signal. 
The crucial point of the ``matrix'' approach is a careful choice of the time interval, 
where the correlator matrix of dimension $N$ has to be dominated by the first $N$ states.

We point out that a precise determination of excited states on the lattice is 
computationally very demanding on its own, since the size and the mass of the state 
increases with the excitation, which in turn implies the use of large lattice
volumes and fine lattice spacings in order to keep under control systematic errors (see 
Ref.~\cite{sasaki}).

In this paper we show that it is possible to access the first excited state
coupled to a single two-point correlator $G(t)$, defining a new correlator which is 
simply an appropriate combination of $G(t)$ taken at different time slices. The mass of 
the first excited state can be extracted from the plateau of the effective mass of the new 
correlator. Therefore our method allows to establish unambiguously the time interval where 
the ground and the first excited states dominate the two-point correlator $G(t)$. In this 
way no prior is needed at all and the extracted resonance mass is not affected by 
contaminations from higher excited states. 

Our procedure is explained in Section~\ref{sec:method}, while its application to 
the extraction of the Roper mass, using the local interpolating field $\mc{O}$, is 
presented in Section~\ref{sec:roper}. In Section~\ref{sec:conclusions}
we draw our conclusions and outlook.

\section{\protect The method \label{sec:method}}

We limit ourselves to consider a single local operator which has a good overlap
with both positive- and negative-parity baryon states, i.e.
\be
\mc{O}_\gamma (x) = \ve_{abc}[u^T_a C \gamma_5 d_b]~u_{c \gamma}(x)~,
\label{op}
\ee
where Latin indices denote color and Greek indices refer to the Dirac
structure, that is contracted in the (ud) diquark on the r.h.s.~of Eq.~(\ref{op}).
Let us then define in the usual way the two-point (zero-momentum) correlation function 
for the operator~(\ref{op}) as
\be
G_{\gamma' \gamma}(t) = \sum_{\mbf{x}} \<0|T\left\{\mc{O}_{\gamma'}(\mbf{x},t) 
\bar{\mc{O}}_{\gamma}(\mbf{0},0)  \right\} |0\>  ~.
\label{2p}
\ee
Adopting periodic boundary conditions on the lattice, Eq.~(\ref{2p}) can be expanded as
\be
G(t) & = & \left( \frac{1 + \gamma_4}{2} \right) \sum_{i \geq 0} 
\left( Z_i^+ e^{-a M_i^+ t} - Z_i^- e^{-a M_i^- (T-t)} \right) \nn \\
 & + & \left( \frac{1 - \gamma_4}{2} \right) \sum_{i \geq 0} 
\left( Z_i^+ e^{-a M_i^+ (T-t)} - Z_i^- e^{-a M_i^- t} \right), \nn \\
\label{2pexp}
\ee
where $t$ is an integer number in units of the lattice spacing $a$, $T$ represents 
the time extension of the lattice, the suffixes $\pm$ refer to the positive- and 
negative-parity states coupled to (\ref{op}), respectively, $M_i^{\pm}$ 
refers to the masses of the given-parity tower of states, and $Z_i^{\pm}$ are the 
corresponding couplings with the interpolating operator.

Choosing $\gamma_4$ diagonal, one gets
\be
G(t) & \equiv & {1 \over 4} \left[ G_{11}(t) + G_{22}(t) + G_{33}(T - t) + G_{44}(T - t) 
\right] \nn \\
& = & \sum_{i \geq 0} \left[ Z_i^+ e^{-a M_i^+ t} - Z_i^- e^{-a M_i^- (T - t)} \right]
\label{2p1st}
\ee
 
Assuming ($T - t$) sufficiently large one has
\be
G(t) = \sum_{i \geq 0} Z_i^+ ~ e^{-a M_i^+ t}
\label{Gt}
\ee 
from which the mass of the positive-parity ground state can be extracted from the 
large-time behavior through an effective mass analysis.

Let us define the following quantity
\be
\widehat{G}(t) \equiv G(t+1) G(t-1) - G^2(t)~,
\label{defhGt}
\ee
which, by direct substitution of Eq.~(\ref{Gt}), can be expanded as
\be
\widehat{G}(t) & = & 2 \sum_{j>i=0}^{\infty} Z_i^+ Z_j^+ e^{-(aM_i^+ + aM_j^+)t} 
\nn \\ & \cdot & \left[ \cosh(aM_j^+ - aM_i^+) - 1 \right] ~ .
\label{hGt}
\ee
One immediately sees that the diagonal terms $i = j$ are cancelled out in the 
new correlator and that the leading exponential decay in Eq.~(\ref{hGt}) is
governed by the \emph{sum} of the masses of the first two states $aM_0^+$ and $aM_1^+$. 
Thus, an effective mass analysis applied to $\widehat{G}(t)$ provides the quantity 
$a (M_0^+ + M_1^+)$. The use of the ``modified'' correlator (\ref{hGt}) is expected to work 
when the mass gap between the ground state and the first excited state is sufficiently large, 
since for a small mass gap the factor $[\cosh(aM_1^+ - aM_0^+) - 1]$, which multiplies 
the exponential in the r.h.s.~of Eq.~(\ref{hGt}), suppresses the signal. 

One can thus set up the following general procedure for extracting the first excited 
state from correlators like~(\ref{Gt}):
\begin{itemize}
\item[i)] Extract the mass of the ground state $aM_0^+$ from the plateau of the effective 
mass for the correlator~(\ref{Gt}). Let the plateau region be given by the time interval
$[t_i, t_f]$.
\item[ii)] Extract the \emph{sum} of the masses of the first two states, $a (M_0^+ + M_1^+)$, 
from the plateau of the effective mass for the``modified" correlator~(\ref{defhGt}). 
Let the plateau region be the time interval $[t^*_i, t^*_f]$. This second plateau is always 
such that $[t^*_i, t^*_f] < [t_i, t_f]$. Indeed, in the time interval $[t_i, t_f]$ the 
correlator~(\ref{Gt}) is dominated by the ground state signal and the modified 
correlator~(\ref{defhGt}) is pure noise, whereas in the time interval $[t^*_i, t^*_f]$ the 
signal associated to the first excited state is visible in both correlators. This in turn means 
that in the larger time interval $[t^*_i, t_f]$ the correlator~(\ref{Gt}) is dominated by the 
first two terms due to the ground and the first excited states, while the contribution from 
higher excited states is negligible, i.e.~within statistical fluctuations.

\item[iii)] Check the consistency of the results obtained from i) and ii) for $aM_0^+$ and 
$aM_1^+$ with those coming from a double-exponential fit of the correlator~(\ref{Gt}) 
in the time interval from $t^*_i$ to $t_f$. 
\end{itemize}

Note that one could in principle consider the modified propagator $\widehat{G}(t)$ 
divided by $G^2(t)$, because such a ratio may directly provide the mass difference 
$a (M_1^+ - M_0^+)$. However the leading exponential decays in $\widehat{G}(t)$ and $G^2(t)$ 
occur in different time intervals, as we will illustrate later, causing a mismatch between 
the plateau of the effective mass of the numerator and the one of the denominator.

In the next Section we apply our method to the extraction of the first excited state 
from the correlator~(\ref{Gt}), and we show that our extracted state can be reasonably 
identified with the Roper resonance\footnote{We mention that our method has been 
successfully applied also to the analysis of the two-point correlator of pseudo-scalar 
mesons. It turns out that the extracted first excited state can be identified with 
$\pi(1300)$ resonance. However, because of both paper's length limitations and the larger 
phenomenological interest of the $N'(1440)$ resonance, we will present in this letter 
only the baryonic results.}. 

\section{\protect Lattice study of the Roper mass \label{sec:roper}}

We have adopted Wilson fermions with non-perturbative Clover improvement and we have generated 450 
quenched gauge configurations at $\beta = 6.20$ in a lattice volume of $24^3 \times 64$.
The hopping parameter is assigned the following values: $k \in \{ 0.1352, 0.1347,
0.1342, 0.1337, 0.1332 \}$. The analysis of pseudoscalar (PS) and vector meson masses 
yields a critical hopping parameter equal to $k_c = 0.13588(2)$ and an inverse 
lattice spacing given by $a^{-1} = 2.55(3)$ GeV. Note that our lattice spacing is 
significantly smaller than the ones adopted in Refs.~\cite{dong,sasaki,leinweber,sbo,bgr1,bgr2}.
The statistical errors for all the quantities considered in this work are evaluated using 
the jacknife grouping procedure \cite{jacknife}. The values of the bare quark mass 
$m_q \equiv (1/k - 1/k_c) / 2a$ and of the PS meson mass are: $m_q \in \{47, 82, 117, 153, 
189 \}$ MeV and $M_{PS} \in \{0.513, 0.686, 0.831, 0.960, 1.08 \}$ GeV.

\subsection{Determination of the positive-parity masses}

The masses of the ground and first excited positive-parity states can be obtained from the 
correlators $G(t)$ [Eq.~(\ref{Gt})] and $\widehat{G}(t)$ [Eq.~(\ref{hGt})]. The ground state 
mass is extracted in a standard way by determining the plateau of the effective mass function 
corresponding to the correlator $G(t)$
\be
a M_{eff}(t) = \ln \left\{ G(t)/G(t+1) \right\}~.
\label{Meff}
\ee
Such a mass is identified with that of the nucleon, and in the following
indicated with $a M_N$.

\begin{figure}[bt]
\includegraphics[scale=0.445]{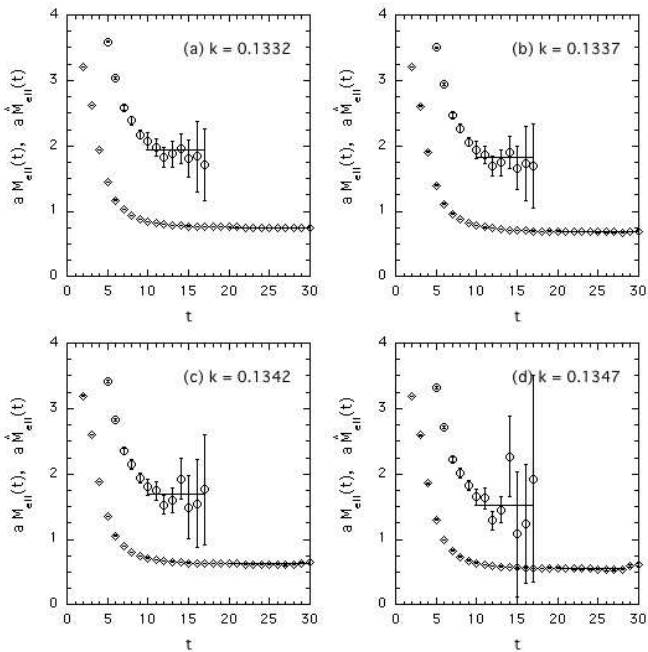}
\caption{Plateaux of the effective mass functions (\ref{Meff}) (diamonds) and (\ref{hMeff}) 
(dots) for $k \in \{ 0.1332, 0.1337, 0.1342, 0.1347 \}$. The solid lines are the 
time extensions considered for the plateaux of Eqs.~(\ref{Meff}) and (\ref{hMeff})}
\label{fig:plateaux}
\end{figure}

The sum of the masses $a M_N + a M_{N'}$ of the ground and first excited states is 
correspondingly obtained from the plateau of the effective mass function related to the 
correlator $\widehat{G}(t)$
\be
a \widehat{M}_{eff}(t) = \ln \left\{ \widehat{G}(t)/\widehat{G}(t+1) \right\}~.
\label{hMeff}
\ee
The quality of the plateaux for $a M_{eff}(t)$ and $a \widehat{M}_{eff}(t)$ is illustrated 
in Fig.~\ref{fig:plateaux}.

In Table~\ref{tab:Meff} we report the mass values $a M_N$ and $a M_{N'}$ obtained from the 
plateaux of Eqs.~(\ref{Meff}) and (\ref{hMeff}), respectively, at the different values of the 
hopping parameter used in the simulations. The time regions of the plateaux are reported in 
Table~\ref{tab:Meff} and explicitly shown in Fig.~\ref{fig:plateaux} by the solid lines. 
Notice that the plateaux of Eq.~(\ref{hMeff}) are of comparable extension with respect to 
those of the ground state effective mass, Eq~(\ref{Meff}).

\begin{table}[htb]
\begin{center}
\begin{tabular}{||c||c||c||c||} 
\hline
\cc $k$ & \cc $\begin{array}{c} t^*_i~-~t^*_f \\ t_i~-~t_f \end{array}$ &
\cc $\begin{array}{c} a M_{N'} \\ a M_N \end{array}$ & $M_{N'} / M_N$ \\

\hline \hline

\cc $0.1332$ & \cc $\begin{array}{c} 10~-~17 \\ 20~-~30 \end{array}$ & 
\cc $\begin{array}{c} 1.189 \pm 0.0181 \\ 0.7534 \pm 0.0027 \end{array}$ & 
$1.578 \pm 0.025$ \\

\hline

\cc $0.1337$ & \cc $\begin{array}{c} 10~-~17 \\ 20~-~30 \end{array}$ & 
\cc $\begin{array}{c} 1.132 \pm 0.022 \\ 0.6911 \pm 0.0033 \end{array}$ & 
$1.638 \pm 0.034$ \\

\hline

\cc $0.1342$ & \cc $\begin{array}{c} 10~-~17 \\ 20~-~30 \end{array}$ &
\cc $\begin{array}{c} 1.067 \pm 0.031 \\ 0.6262 \pm 0.0042 \end{array}$ & 
$1.705 \pm 0.052$ \\

\hline

\cc $0.1347$ & \cc $\begin{array}{c} 10~-~17 \\ 20~-~28 \end{array}$ & 
\cc $\begin{array}{c} 0.969 \pm 0.039 \\ 0.5561 \pm 0.0058 \end{array}$ & 
$1.743 \pm 0.075$ \\

\hline

\cc $0.1352$ & \cc $\begin{array}{c} 10~-~17 \\ 20~-~28 \end{array}$ & 
\cc $\begin{array}{c} 0.825 \pm 0.053 \\ 0.4832 \pm 0.0054 \end{array}$ & 
$1.71 \pm 0.11$ \\

\hline
\end{tabular}
\caption{Effective masses $aM_N$ and $aM_{N'}$ at the different values of $k$ used
in the simulations. The second column reports the time extensions of the
plateaux of Eqs.~(\ref{Meff}) and (\ref{hMeff}).}
\label{tab:Meff}
\end{center}
\end{table}
 
\begin{figure}[bt]
\includegraphics[scale=0.445]{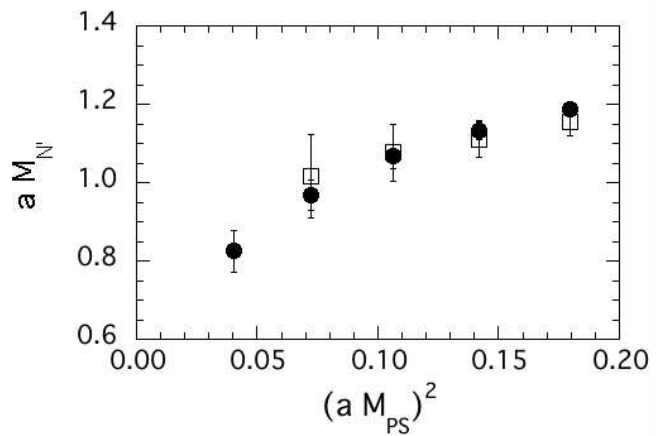}
\caption{Comparison of the mass of the first excited positive-parity state, $a M_{N'}$, 
obtained for different quark masses using the effective mass method (full dots) and the 
double-exponential fit (open squares).}
\label{fig:comparison}
\end{figure}

To check the consistency of our results, we have then performed a
double-exponential fit of the correlator~(\ref{Gt}) in the time coordinate, 
limiting the fitting region to the one including the plateaux of the two effective 
mass analyses, i.e. $[t^*_i, t_f]$. In this fit we have considered two options for 
the nucleon mass $aM_N$: a) a free parameter to be fixed by the fitting 
procedure; b) a parameter fixed at the value obtained from the effective mass analysis 
of Eq.~(\ref{Meff}) [see the third column of Table~\ref{tab:Meff}]. The two options provide 
consistent results for both $aM_N$ and $aM_{N'}$, but the uncertainties are definitely 
smaller in the case b). Our results for $aM_{N'}$, corresponding to the latter option, are 
reported in Table~\ref{tab:2exp}. The consistency between the results of the effective mass 
analysis and the double-exponential fit can be appreciated from Fig.~\ref{fig:comparison}, where it can 
be seen that the more precise results are those obtained from the effective mass analysis.

\begin{table}[htb]
\begin{center}
\begin{tabular}{||c||c||c||} 
\hline
\cc $k$ & \cc $t^*_i~-~t_f$ & $a M_{N'}$\\

\hline \hline

\cc $0.1332$ & \cc $10~-~30$ & $1.154 \pm 0.033$ \\
\hline
\cc $0.1337$ & \cc $10~-~30$ & $1.111 \pm 0.047$ \\
\hline
\cc $0.1342$ & \cc $10~-~30$ & $1.076 \pm 0.074$ \\
\hline
\cc $0.1347$ & \cc $10~-~28$ & $1.02 \pm 0.11$ \\
\hline
\cc $0.1352$ & \cc $10~-~28$ & -- \\

\hline
\end{tabular}
\caption{Results for the mass of the first excited positive-parity state obtained from a
double-exponential fit of the correlator~(\ref{Gt}) in the time region $[t_i^*, t_f]$ which 
includes the plateaux of the effective masses (\ref{Meff}) and (\ref{hMeff}). In this fit the 
nucleon mass $aM_N$ is constrained to be the one obtained from the effective mass analysis 
(\ref{Meff}).}
\label{tab:2exp}
\end{center}
\end{table}

\subsection{Determination of the negative-parity mass $M_{N^*}$}

The analysis of the negative-parity ground state of the correlator~(\ref{Gt}) strictly
follows that of the positive-parity one, described in the previous subsection. One simply
has to consider $t$ sufficiently large so that Eq.~(\ref{Gt}) can be replaced by
\be
G(t) = - \sum_{i \geq 0} Z_i^- ~ e^{-a M_i^- (T - t)}
\label{GTt}
\ee 

\begin{table}[htb]

\begin{center}
\begin{tabular}{||c||c||c||}
\hline
$k$ & $(T - t)_i~-~(T - t)_f$ & $a M_{N^*}$\\

\hline \hline

\cc $0.1332$ & \cc $14~-~19$ & $0.926 \pm 0.009$ \\
\hline
\cc $0.1337$ & \cc $14~-~19$ & $0.860 \pm 0.011$ \\
\hline
\cc $0.1342$ & \cc $14~-~19$ & $0.791 \pm 0.015$ \\
\hline
\cc $0.1347$ & \cc $14~-~19$ & $0.717 \pm 0.023$ \\
\hline
\cc $0.1352$ & \cc $12~-~17$ & $0.697 \pm 0.026$ \\

\hline
\end{tabular}
\caption{Results for the negative-parity ground-state mass $a M_{N^*}$ obtained through the 
effective mass analysis (\ref{Meff}). The time intervals of the plateaux are also reported.}
\label{tab:s11}
\end{center}

\end{table}

\begin{figure}[htb]
\includegraphics[scale=0.445]{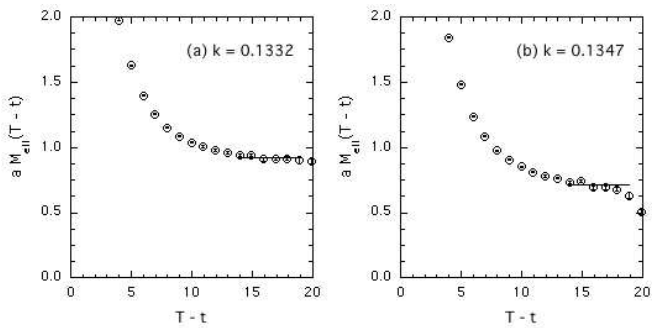}
\caption{Plateaux of the effective mass function (\ref{Meff}) for the (negative-parity) 
correlator (\ref{GTt}) for $ k = 0.1332$ (a) and $k = 0.1347$ (b).}
\label{fig:plateaux_s11}
\end{figure}

In Table~\ref{tab:s11} we report the values obtained for $aM_{N^*}$ from the effective mass 
analysis of the correlator (\ref{GTt}), together with the corresponding time intervals for 
the plateaux. The latter are shown in Fig.~\ref{fig:plateaux_s11} for a couple of values of 
the quark mass. It can be seen that the ground-state signal is of good quality, but with 
respect to the case of positive-parity the plateaux are narrower and shifted to lower values 
of the lattice time, mainly because of the higher mass of the negative-parity ground state.

\begin{figure}[htb]
\includegraphics[scale=0.445]{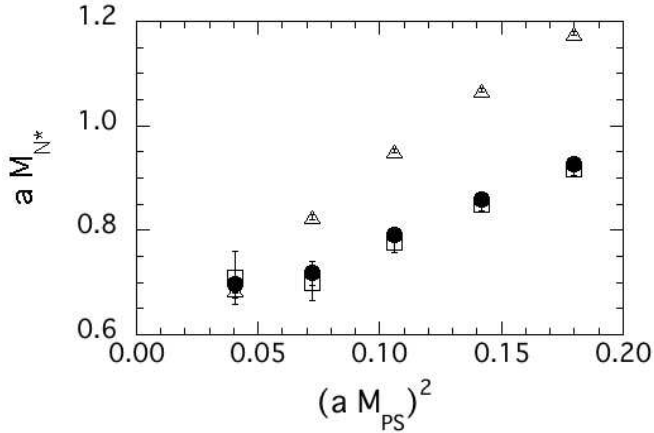}
\caption{Results obtained for $aM_{N^*}$ from the effective mass analysis (full dots) and 
from a single-exponential fit in the time interval (open squares) versus 
the square of the PS mass. The sum of the masses $aM_N + aM_{PS}$ (open triangles) 
is also plotted for comparison.}
\label{fig:s11}
\end{figure}

In Fig.~\ref{fig:s11} our results for $aM_{N^*}$ are plotted versus the squared PS 
mass $(a M_{PS})^2$. As a consistency check, we have performed a single-exponential fit of 
the correlator (\ref{GTt}) in the time interval corresponding to the plateaux reported in 
Table~\ref{tab:s11}. The results are shown in Fig.~\ref{fig:s11} as open squares.

Our lattice simulations can be directly compared with those of Ref.~\cite{goeckeler}, since 
the only difference is the use of smeared interpolating operators in the latter. The two 
simulations nicely agree in the results for both $a M_N$ and $a M_{N^*}$.

In Fig.~\ref{fig:s11} the mass $a M_N + a M_{PS}$ of the (S-wave) scattering state 
$N + PS$ at zero hadron momenta, is reported for comparison. Note that such a scattering 
state stays higher in energy with respect to the resonance state for the quark masses used 
in the present simulations, except for the lightest mass where they nearly coincide (cf.~also 
Ref.~\cite{goeckeler}). In principle the mass of the scattering state can be extracted using 
the modified correlator (\ref{hGt}) constructed as in Eq.~(\ref{defhGt}) from the 
negative-parity propagator (\ref{GTt}). However in practice, due to the occurrence of the 
plateaux of the negative-parity ground-state at quite earlier times compared with the case 
of the positive-parity one (see Tables~\ref{tab:Meff} and \ref{tab:s11}), as well as to a 
smaller mass gap and to larger statistical fluctuations, we find no clear signal of the 
scattering state $N + PS$ in the modified correlator.

Though it is beyond the aim of the present study, we want to briefly comment on the 
systematic errors of our simulations before closing the subsection.

Our calculations have been carried out at a fixed value of the lattice spacing $a$. The use 
of the non-perturbative Clover improvement guarantees that discretization errors of order 
$\mc{O}(a)$ are absent in our extracted masses. Nevertheless the closeness to the continuum 
limit strongly depends on the condition $a M \ll 1$, which is better fulfilled by our results 
for the positive-parity ground-state (the nucleon). In case of $a M_{N^*}$ and $a M_{N'}$ 
the discretization error should be investigated by performing simulations at different values 
of the lattice spacing. In this respect we point out again that the value of $a$ employed in 
our study is much lower than those used in most of the works appeared in the literature on 
the subject, like Refs.~\cite{dong,sasaki,leinweber,sbo,bgr1,bgr2}.

As shown in \cite{sasaki}, finite volume effects can affect the extracted mass of excited 
states. The estimates made in \cite{sasaki} indicate that at the volume and the quark 
masses used in our lattice simulations one can expect a $\approx 10 \div 20\%$ reduction of 
the resonance masses. 

Finally, the error due to the quenched approximation can be estimated only by removing such 
approximation. Available results in the literature \cite{ground} suggest that the quenching 
error on the ground-sate baryon spectrum is not large and expected within $\approx 10 \%$. 
However, very little is known about the impact of the quenched approximation on the excited 
baryon spectrum.

\subsection{Extrapolation to the physical point}

\begin{figure}[htb]
\includegraphics[scale=0.445]{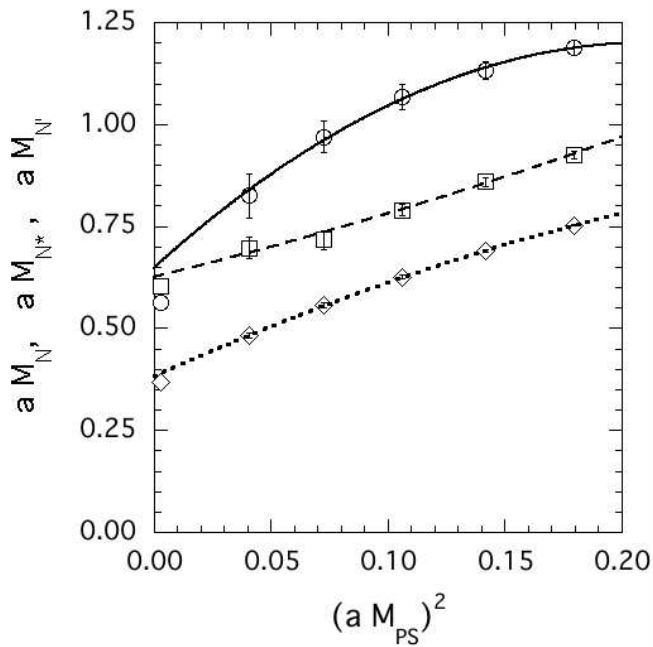}
\caption{Plot of $a M_N$ (diamonds), $a M_{N^*}$ (squares) and $a M_{N'}$ (dots) versus 
the square of the PS mass. The dotted, dashed and solid lines are analytical fits 
of the form $A + B (a M_{PS})^2 + C (a M_{PS})^4$. The markers at the physical point 
correspond in lattice units to the physical masses of the nucleon N(938), the $S_{11}(1535)$ 
and $N'(1440)$ resonances.}
\label{fig:Nucleon+Roper+s11}
\end{figure}

All our lattice results for $a M_N$, $a M_{N^*}$ and $a M_{N'}$ are shown in 
Fig.~\ref{fig:Nucleon+Roper+s11} versus the square of the PS mass. Up to the lightest 
quark mass used in our simulations there is no level switching between the negative-parity 
ground-state and the positive-parity first excited state, the latter lying above the former 
at variance with the physical spectrum.

The extrapolation to the physical point should be guided by chiral perturbation theory. 
However, the quark masses used in our simulations are quite far from the chiral regime and 
indeed we expect that the results shown in Fig.~\ref{fig:Nucleon+Roper+s11} do not contain 
sizable effects from chiral logs. In order to identify the masses found on the lattice with 
the resonances of the physical spectrum, one should perform simulations at lower quark masses. 
For a na\"ive correspondence we will limit ourselves to adopt an analytical quark mass 
dependence of the form $a M = A + B (a M_{PS})^2 + C (a M_{PS})^4$, which is represented 
by the lines in Fig.~\ref{fig:Nucleon+Roper+s11} for the various states. The positive- 
and negative-parity ground states can be reasonably identified with nucleon $N(938)$ and 
the $S_{11}(1535)$ resonance. As for the positive-parity first excited state, the quark mass 
dependence of our extracted masses is consistent with the identification with the Roper 
resonance $N'(1440)$, though a possible interpretation as the higher nucleon resonance 
$N(1710)$ cannot be completely ruled out. Such a possible extrapolation ambiguity is 
not addressed in the literature~\cite{dong,sasaki,leinweber,sbo,bgr1}, and a clear-cut 
solution again relies in performing lattice simulations at lower quark masses. The latter 
would also help clarifying at which quark masses the level switching between the orbital 
and radial excitations of the nucleon occurs. Basing on our crude extrapolations such a 
crossing is tentatively expected to appear close to the physical point.

\section{\protect Conclusions \label{sec:conclusions}}

We have presented a new method for extracting excited states from a single two-point
correlation function calculated on the lattice. We have defined a new correlator which 
is simply an appropriate combination of the correlation function evaluated at different 
time slices. In this way the leading exponential decay (ground state) is exactly subtracted 
and one can access the first excited state. The mass of this state can be extracted from the 
plateau of the effective mass of the new correlator. Therefore our method allows to establish 
unambiguously the time interval where the ground and the first excited states dominate the 
two-point correlation function. 

Our method has been applied to a quenched lattice study (volume = $24^3 \times 64$, 
$\beta = 6.2$, $a^{-1} = 2.55$ GeV) of the first excited state of the nucleon 
using the interpolating operator $\mc{O} = \ve_{abc}[u^T_a C \gamma_5 d_b]u_c$.
The results are consistent with the identification of our extracted excited 
state with the Roper resonance $N'(1440)$. The switching of the level ordering 
with respect to the negative-parity partner of the nucleon, $N^*(1535)$, is 
not seen at the simulated quark masses and, basing on crude extrapolations, is 
tentatively expected to occur close to the physical point.

\section*{Acknowledgements} We thank G.~Martinelli for useful comments and 
a careful reading of the manuscript. We thank also G.T.~Fleming, C.~Gattringer and N.~Mathur 
for useful correspondence.

\end{document}